\documentclass[english, 12pt, letterpaper]{article}
\usepackage{blindtext}
\usepackage{amsfonts, amssymb, ejgAms, babel,
wscomment, bbm, 
layouts, hyperref, natbib, tikz
}
\newcommand{\uu}{u(x_{i1},  x_{i2})}
\newcommand{\vv}{u(x_{i1}, 0)}
\newcommand{\ww}{u(0, x_{i2})}

\newcommand{\x}{(x_{i1}, x_{i2})}
\newcommand{\bb}{(b_{i1}, b_{i2})}
\newcommand{\qb}{Q_1(b_{i1})}
\newcommand{\qbb}{Q_2(b_{i2})}
\newcommand{\ub}{\vv-b_{i1}}
\newcommand{\ubb}{\ww-b_{i2}}
\newcommand{\bh}{b_{i1}^h}
\newcommand{\bbh}{b_{i2}^h}
\newcommand{\bl}{b_{i1}^l}
\newcommand{\bbl}{b_{i2}^l}
\newcommand{\qbh}{Q_1(\bh)}
\newcommand{\qbbh}{Q_2(\bbh)}
\newcommand{\qbl}{Q_1(\bl)}
\newcommand{\qbbl}{Q_2(\bbl)}
\newcommand{\xh}{x_{i1}^h}
\newcommand{\xxh}{x_{i2}^h}
\newcommand{\xl}{x_{i1}^l}
\newcommand{\xxl}{x_{i2}^l}
\newcommand{\uul}{u(\xl, \xxl)}
\newcommand{\vvl}{u(\xl,0)}
\newcommand{\wwl}{u(0, \xxl)}

\newcommand{\vvh}{u(\xh, 0)}
\newcommand{\wwh}{u(0, \xxh)}
\newcommand{\smi}{s_{-i}}
\newcommand{\mus}{\mu_{\smi}}

\makeatletter
\def\blfootnote{\xdef\@thefnmark{}\@footnotetext}
\makeatother

\begin{document}

\title{Simultaneous Auctions for Complementary Goods}
\author{Wiroy Shin}

 \smallskip 
\date{December 9, 2013
\blfootnote{First draft : August 15, 2012}}

\maketitle

\begin{abstract}
This paper studies an environment of simultaneous, separate, first-price auctions for complementary goods. 
Agents observe private values of each good before making bids, and the complementarity between goods  is explicitly incorporated in their utility. For simplicity, a model is presented with two first-price auctions and two bidders. We show that a monotone pure-strategy Bayesian Nash Equilibrium exists in the environment. 

\end{abstract}

\blfootnote{Department of Economics, The Pennsylvania State University, University Park, PA, 16802, USA.}
\blfootnote{\textit{E-mail address}: \href{mailto:wus130@psu.edu}{\texttt{wus130@psu.edu}}}
\blfootnote{I thank to Edward J. Green for his guidance and support throughout this project. I also gratefully acknowledge the comments of James Jordan, Leslie Marx, Ed Coulson, and Paul Grieco. All errors are mine.}

\section{Introduction}

This paper studies an environment of simultaneous, separate, first-price auctions for complementary goods. Agents observe private values of each good before making bids, and the complementarity between goods  is explicitly incorporated in their utility. For simplicity, a model is presented with two first-price auctions and two bidders. We show that a monotone pure-strategy Bayesian Nash Equilibrium exists in the environment. This model, in which the auction mechanism has an exogenous (albeit
widely used) form and in which sellers do not interact strategically
with one another, is intended to provide a tractable framework for
applied research focusing on competition among buyers. Each buyer is
trying to acquire a portfolio of objects that must be purchased by
auction from several unrelated sellers. eBay's various sealed-bid auctions under same collectibles category which are run by different sellers with similar closing time can be an example of this environment.  Also, less explicit, but equally compelling, auctions for acquisition of firms manufacturing a final good and an intermediate good input can be another example. Here, a buyer intends to create a vertically integrated firm and the two acquired firms can be viewed as complements when they are merged.  \\

There are several branches of existing auction research that are similar
to, but distinct from, what is being done here. Let us comment briefly
on those contributions, and on how the present research is distinguished
from them.
A literature on multi-unit auctions, including \cite{bresky1999equilibria} and \cite{mcadams2003isotone}, studies sale of multiple identical goods in an auction with specific pricing rules. As modeled, multiple identical objects resemble substitutable distinct objects rather than complementary ones modeled here. Package auctions(see \cite{bernheim1986menu} and \cite{ausubel2002ascending}) model multiple goods without assuming substitutability only. They suppose that an explicit bid can be offered for a package of objects, which is not possible when various objects are auctioned by distinct and unrelated sellers.\\

There are a few prior research contributions on simultaneous auctions, including \cite{krishna1996simultaneous} and \cite{bikhchandani1999auctions}. They either assume complete information on bidders' valuations or employ a single parameter for private value signals of multiple objects. However, such parameterization is too rigid to model heterogeneity of bidders' preferences on each item and complementarity.
In this paper, we relax those restrictions so that the environment in which each bidder receives private independent signals for each heterogeneous object is permitted.\\

We follow \cite{reny2011existence} to prove the existence of a monotone pure-strategy equilibrium. Along with basic assumptions on the Bayesian games(G.1-G.6, p.509  \cite{reny2011existence}), \emph{weak single crossing property} and \emph{weak quasi supermodularity} are sufficient for the existence theorem in his paper.\footnote{\cite{reny2011existence}, Theorem 4.1 and Proposition 4.4 } \\ 


The remainder of this paper as follows. In the next section, we present an environment of simultaneous separate complementary first-price auctions in detail, and proposes a theorem for the existence of a monotone pure-strategy equilibrium. Section 3 provides the proof, and section 4 discusses an extension of the study to a continuum bid space setting.


\section{Simultaneous first-price auctions for complementary goods}

In this section, we define simultaneous first-price auctions for complementary goods with 2 non-identical objects and 2 bidders. This specific auction environment is denoted by $\mathbb{M}$. At the end of the section, we propose a theorem for the existence of monotone pure-strategy equilibrium. \\

\subsection{Environment}

\begin{description}

\item[Bidder] $i\in\{1,2\}$.

\item[Object] $k\in\{1,2\}$, Values of objects\footnote{It can be interpreted as a type of an agent.} are represented by vector ${x_{i}}=(x_{i1},\, x_{i2})$. $x_{ik}\sim F(\cdot)\,$ and$\ x_{ik}\in X=[0,1] $. Let $F(\cdot)$ be atomless.\\
; $x_{ik}$ is a random variable with a CDF $\ F(\cdot),$ which has a support $X$ and a PDF $f:X \to [0,1]$. 
Each object $k$ is distributed $i.i.d.$ Therefore, joint CDF and joint PDF for two objects $(x_{i1},\, x_{i2})$ are given by $G(x_{i1},\, x_{i2})=  F(x_{i1})\times F(x_{i2})$,  and $g(x_{i1},\, x_{i2})=f(x_{i1})\times f(x_{i2})$, respectively.


\item[Bid] ${b_{i}}=(b_{i1},\, b_{i2})\in B\times B = A \subset \Re^2$. Each bidder $i$ places bids $b_{i1}$ and $ b_{i2}$ simultaneously. $B$ is a finite subset of $\Re$. That is, we assume that the action space of this game, $A$,  is finite.

\item[Utility] Each bidder has a quasi linear ex-post utility $U_i(\cdot, \cdot, \cdot, \cdot)$. 
\disparray a{
U_{i}({b_{i}},\,{b_{j}},\,{x_{i}},\,{x_{j}})=\begin{cases}
u(x_{i1}, x_{i2})-(b_{i1}+b_{i2}) & if\ b_{i1}> b_{j1}\, and\, b_{i2} > b_{j2}\ ,where\ i\neq j\\
u(x_{i1}, 0)-b_{i1} & if\ b_{i1} > b_{j1}\, and\, b_{i2} <b_{j2}\\
u(0, x_{i2})-b_{i2} & if\ b_{i1}<b_{j1}\, and\, b_{i2} > b_{j2}\\
0 & \, otherwise.
\end{cases}}

Note that when ties occur the object is allocated randomly to one of two bidders with $\frac{1}{2}$ probablity.
The ex-post utility can be interpreted as follows. An agent realizes his utility from objects he wins, through a function $u: \Re^2 \rightarrow \Re$. $u(\cdot, \cdot)$ transforms the independent values of goods to utility which counts  a synergy effect between two objects, \emph{e.g.} $\uu= x_{i1} + x_{i2} + \alpha \cdot \mathbbm{1}_{\{x_{ik} > 0,\ k\in \{1,2\}\}}$, where $\alpha >0$. And then, the bidding values for the objects he wins are subtracted from $u(\cdot, \cdot)$.

Here are additional assumptions on the $u(\xi)$.\\

(A1) $u(x_i)$ is increasing in $x_i$. \\ 

(A2) $\lambda \x=\uu-[\vv+\ww] \geq 0 \ , \ where\ x_{i1} \geq 0 \ and \ \ x_{i2} \geq 0 $ \\
; This assumption implies that regardless of values of objects, agents realize that utility from winning both objects is greater than or equal to sum of utilities of each good. (e.g. Goods are complementary, Owning all bike shops in New York City yields monopolistic positioning benefit in the business) \\ 

(A3) Let $\geq$ be a coordinatewise partial order on types. That is, $[x_i\geq y_i \Leftrightarrow (x_{i1} \geq y_{i1}) \ {and}\ (x_{i2} \geq y_{i2}) ]$. Assume $[x_i\geq y_i \Rightarrow \lambda(x_i) \geq \lambda(y_i) ]$. In other words, $\lambda$ is monotone in $x_i=\x.$ \\\

In summary , bidders expect an extra synergy by obtaining both items, and the value of the synergy is weakly increasing with values of obtained goods.

\end{description}

Since it is reasonable not to bid more than the maximum of utility an agent can earn, we restrict the minimum and maximum bid on one good as $u(0,0)$ and $u(1,1)$ respectively. Let  $u(1,1)$ be denoted by $\bar{u}$. Without loss of generality, we impose $u(0,0)=0$ and constant increments on bids. Accordingly, the action space is $A=\{0, \frac{1}{n}, \frac{2}{n}, ..., \bar{u}\} \times \{0, \frac{1}{n}, \frac{2}{n}, ..., \bar{u}\}, n\in \mathbb{Z}_+.$  

\subsection{Monotone pure-strategy equilibrium}

A \textit{pure-strategy} for bidder $i$ is a function $s_i:X^2 \rightarrow A$ and it is \textit{monotone} if $x'_i \geq x_i$ implies $ s_i(x'_i) \geq s_i(x_i)$, for all $x'_i ,x_i \in X^2$. Let $S_i$ be a set of monotone pure strategies of $i$. Based on the environment in section 2.1, bidder $i$'s expected utility can be written as follows.\\

Let  $\mus$ be a distribution of an opponent's bids constructed by $s_{-i}(x_{-i})$ and $G(x_{-i})$. Define
\disparray d{
P_3(b_i) &=& \mu_{s_{-i}}([0, b_{i1}) \times [0, b_{i2})) + \frac{1}{2}\mu_{s_{-i}}([ b_{i1}, b_{i1}] \times [0, b_{i2})) \\&&+ \frac{1}{2}\mu_{s_{-i}}([0, b_{i1}) \times [b_{i2}, b_{i2}])+\frac{1}{4} \mu_{s_{-i}}([ b_{i1}, b_{i1}]  \times [b_{i2}, b_{i2}]); \\
P_1(b_i) &=& \mu_{s_{-i}}([0, b_{i1}) \times (b_{i2}, \bar{u}]) + \frac{1}{2}\mu_{s_{-i}}([ 0, b_{i1}) \times [b_{i2}, b_{i2}]) \\&&+ \frac{1}{2}\mu_{s_{-i}}([b_{i1}, b_{i1}] \times (b_{i2}, \bar{u} ]) +\frac{1}{4} \mu_{s_{-i}}([ b_{i1}, b_{i1}]  \times [b_{i2}, b_{i2}]); \\
P_2(b_i) &=& \mu_{s_{-i}}((b_{i1}, \bar{u}) \times [0, b_{i2})) + \frac{1}{2}\mu_{s_{-i}}([b_{i1}, b_{i1}] \times [0, b_{i2})) \\&&+ \frac{1}{2}\mu_{s_{-i}}((b_{i1}, \bar{u}] \times [b_{i2}, b_{i2} ]) +\frac{1}{4} \mu_{s_{-i}}([ b_{i1}, b_{i1}]  \times [b_{i2}, b_{i2}]).
}

$P_3(b_i)$ denotes a probability that bidder $i$ wins both objects with $b_i$ when the opponent plays $s_{-i}$. $P_1(b_i)$ and $P_2(b_i)$ indicate probabilities for winning object 1 \emph{only} and object 2 \emph{only}, respectively. Then, if an agent with type $x_i$ bids $b_i$, the bidder $i$'s expected utility is
\display e{
V_i(b_i,x_i, s_{-i}) = P_3(b_i)[u(x_{i1}, x_{i2}) - (b_{i1} + b_{i2})] +
P_1(b_i)[u(x_{i1}, 0)-b_{i1}] + P_2(b_i)[u(0, x_{i2})-b_{i2}].}

To make proofs easy in section 3, we modify a representation of the expected utility. Define
\disparray f{
Q_3 \bb &=& P_3(b_i);\\
Q_1\bb &=& \mus([0, b_{i1})\times [0, \bar{u}]) + \frac{1}{2}  \mus([b_{i1}, b_{i1}]\times [0, \bar{u}]) ;\\
Q_2\bb &=& \mus([0, \bar{u}] \times [0, b_{i2})) + \frac{1}{2}  \mus([0, \bar{u}] \times [b_{i2}, b_{i2}]) .
}

$Q_1\bb$ denotes a probability that bidder $i$ with a bid $b_{i1}$ wins the object 1 at least. $Q_2\bb$ also describes a probablity that bidder $i$ with a bid $b_{i2}$ wins the object 2 at least. Note that $Q_1$ only depends on $b_{i1}$ and is increasing in $b_{i1}$, and $Q_2$ only depends on $b_{i2}$ and is increasing in $b_{i2}$. For notational convenience, we eliminate $b_{i2}$ and $b_{i1}$ in $Q_1$ and $Q_2$'s argument respectively.  $Q_3$ is increasing in both $b_{i1}$ and $b_{i2}$.\\

The way that constructing $Q_1, Q_2$ and $Q_3$ infers [$P_3(b_i)=Q_3(b_i), \ P_1(b_{i1})=Q_1(b_{i1})-Q_3(b_i), \ P_2(b_i)=Q_2(b_{i2})-Q_3(b_i)$]. \\

Then, \eqn e becomes
\disparray g{
V_i(b_i,x_i, s_{-i})&=& \qb[\ub]+\qbb[\ubb]\\&&+Q_3 \bb[\uu-(b_{i1}+b_{i2})-(\ub)-(\ubb)] \\
&=&  \qb[\ub]+\qbb[\ubb]\\&&+Q_3 \bb[\uu-\vv-\ww] \\
&=&  \qb[\ub]+\qbb[\ubb]+Q_3 \bb \lambda(x_i).
}

Under the auction $\mathbb{M}$, $(s_1, s_2)\in S_1\times S_2$ is a \textit{monotone pure-strategy equilibrium}, if for every bidder $i\in \{1,2\}$ and every $x_i\in X^2$, 

\display I{V_i(s_i(x_i),x_i, s_{-i}) \geq V_i(b_i,x_i, s_{-i})} for all $b_i\in A$.

\Theorem{Suppose an auction satisfies the environment $\mathbb{M}$ which is defined in section 2.1. Then, there exist a monotone pure-strategy equilibrium for the auction $\mathbb{M}$.}\\


\section{Proof of Theorem 1}

\textnormal{Under the auction $\mathbb{M, }$ we have an atomless CDF  for object values with a two-dimensional compact support. Therefore, by propositon 3.1 in \cite{reny2011existence}, assumptions G.1-G.5 on the Bayesian game (p.509 \cite{reny2011existence}) are satisfied. Also, with a finite action space $A$ definied in section 3.1, the action space is a lattice, and the ex-post utility $U_i$ is bounded, jointly measurable and continuous in $b_i\in A$ for every $x_i \in X^2$. By the proposition 4.4(i) in \cite{reny2011existence}, if  the interim utility $V_i(b_i,x_i, s_{-i})$ satisfies weak single crossing property and quasi supermodularity, \textbf{Theorem 1} is proved.}

\subsection{Weak single crossing property of $V_i(b_i,x_i, s_{-i})$}

\Definition{$V_i(b_i,x_i, s_{-i})$ satisfies \emph{weak single crossing} if for all monotone pure strategies $s_{-i}$ of the opponent, if for all $b'_i\geq b_i$, and for all $x'_i \geq x_i$, 
\display o{V_i(b'_i,x_i, s_{-i})\geq V_i(b_i,x_i, s_{-i}) \ \text{implies}\ V_i(b'_i,x'_i, s_{-i})\geq V_i(b_i,x'_i, s_{-i}). }
}

Let $b_i'=(\bh, \bbh), \ b_i=(\bl, \bbl),\  x_i'=(\xh, \xxh), \ x_i=(\xl, \xxl).$

\Lemma{With $b_i, b'_i, x_i$ and $x'_i$, \eqn o holds. That is, 
\display p{V_i(\bh, \bbh, x_i, s_{-i})-V_i(\bl, \bbl, x_i, s_{-i}) \geq 0 } implies
\display q{V_i(\bh, \bbh,x'_i, s_{-i})-V_i(\bl, \bbl,x'_i, s_{-i}) \geq 0.}}
\proof{\eqn p gives
\disparray r{&&\qbh[\vvl-\bh]+\qbbh[\wwl-\bbh]\\&&+Q_3(\bh, \bbh)[\uul-\vvl-\wwl]\\
\geq&& \qbl[\vvl-\bl]+\qbbl[\wwl-\bbl]\\&&+Q_3(\bl, \bbl)[\uul-\vvl-\wwl] }
\disparray s{
\Leftrightarrow && [Q_3(\bh, \bbh)-Q_3(\bl, \bbl)] \lambda(x_i) \\ 
\geq&& [\qbl[\vvl-\bl]-\qbh[\vvl-\bh]]\\&&+ [\qbbl[\vvl-\bbl]-\qbbh[\vvl-\bbh]]
}\\
We want to show that \eqn s gives \eqn q. Rewriting \eqn q, 
\disparray t{
 && [Q_3(\bh, \bbh)-Q_3(\bl, \bbl)] \lambda(x_i') \\ 
\geq&& [\qbl[\vvh-\bl]-\qbh[\vvh-\bh]]\\&&+ [\qbbl[\wwh-\bbl]-\qbbh[\wwh-\bbh]]
}\\
To prove lemma, it is sufficient to show that $[(LHS \eqn t \geq LHS \eqn s) \  and\ ( RHS \eqn s \geq RHS \eqn t)]$.
\disparray u{
LHS \eqn t - LHS \eqn s &=& [Q_3(\bh, \bbh)-Q_3(\bl, \bbl)][\lambda(x_i')-\lambda(x_i)] \geq 0.
}
The inequality comes from the fact that both terms are positive. $Q_3$ is increasing in its argument. Also,  $\lambda(x_i)$ is monotone by (A3) in section 2.1, and $x_i' \geq x_i$.

\disparray v{
RHS \eqn s - RHS \eqn t & =&  \qbl[\vvl-\vvh]-\qbh[\vvl-\vvh] \\ &&+\qbbl[\wwl-\wwh]-\qbbh[\wwl-\wwh]\\
&=&[\qbl-\qbh][\vvl-\vvh]\\&&+[\qbbl-\qbbh][\wwl-\wwh]\\
&\geq&0
}
The inequality comes from the fact that $Q_1, Q_2$ and $u$ is increasing in its argument. $\blacklozenge$

}

Therefore, by Lemma 1, $V_i(b_i,x_i, s_{-i})$ satisfies weak single crossing property.


\subsection{Weak quasi-supermodularity of $V_i(b_i,x_i, s_{-i})$}
\smallskip
\Definition $V_i(b_i,x_i, s_{-i})$ is \emph{weakly quasisupermodular} if for all monotone pure strategies $s_{-i}$ of the opponent, if for all $b_i, b'_i\in A$, and every $x_i \in X^2$, 
\display h{
V_i(b_i,x_i, s_{-i}) \geq V_i(b_i\wedge b'_i, x_i, s_{-i}) \ \text{implies} \ V_i(b_i\vee b'_i, x_i, s_{-i})\geq V_i(b'_i,x_i, s_{-i}).
}

If $b'_i\geq b_i$, where $\geq$ is a coordinatewise partial order, \eqn h holds trivially. Therefore, we focus on the case $[(b_{i1}> b_{i1}') \ and \ (b_{i2} < b_{i2}')]$.
Let $b_{ik}^h, \ b_{ik}^l \in B$ and $b_{ik}^h>b_{ik}^l$ for any $k\in\{1,2\}.$ For the current case, let $b_i=(\bh, \bbl)$ and $b'_i=(\bl, \bbh)$.\\

Substitute $b_i, b'_i$, to \eqn h. That is, 
\display i{V_i(\bh, \bbl, x_i, s_{-i})-V_i(\bl, \bbl, x_i, s_{-i}) \geq 0 } implies
\display j{V_i(\bh, \bbh,x_i, s_{-i})-V_i(\bl, \bbh,x_i, s_{-i}) \geq 0.}

\eqn i gives
\disparray k{&&\qbh[\vv-\bh]+\qbbl[\ww-\bbl]\\&&+Q_3(\bh, \bbl)[\uu-\vv-\ww]\\
\geq &&\qbl[\vv-\bl]+\qbbl[\ww-\bbl]\\&&+Q_3(\bl, \bbl)[\uu-\vv-\ww] }
\display l{
\Leftrightarrow \qbh[\vv-\bh]-\qbl[\vv-\bl] \geq [Q_3(\bl, \bbl)-Q_3(\bh, \bbl)]\lambda(x_i).
}\\

Rewriting \eqn j, 
\display m{ \qbh[\vv-\bh]-\qbl[\vv-\bl] \geq [Q_3(\bl, \bbh)-Q_3(\bh, \bbh)]\lambda(x_i).}
Since LHS of \eqn l and \eqn m are identical,  a sufficient condition for weak quasisupermodularity is  [RHS \eqn l $\geq$ RHS \eqn m].

\disparray n{
RHS \eqn l - RHS \eqn m =&&\lambda(x_i) \cdot \\&&[[Q_3(\bh, \bbh)-Q_3(\bh, \bbl)] -[Q_3(\bl, \bbh)-Q_3(\bl, \bbl)]].  
}
With the assumption(A2), $\lambda(x_i)\geq 0$, for weak quasisupermodularity,  it is sufficient to show that  [RHS \eqn l - RHS \eqn m] is positive. That is,

\display w{[Q_3(\bh, \bbh)-Q_3(\bh, \bbl)] -[Q_3(\bl, \bbh)-Q_3(\bl, \bbl)] \geq 0.}\\

\Lemma Recall that in the auction $\mathbb{M}$,  $X^2=[0,1]^2$ and  $A=\{0, \frac{1}{n}, \frac{2}{n}, ..., \bar{u}\} \times \{0, \frac{1}{n}, \frac{2}{n}, ..., \bar{u}\}, n\in \mathbb{Z}_+$.  In this environment, inequality \eqn w holds for all monotone pure-strategies of the opponent.
 \proof{
Let $x_j\in X\times X$ and $p(x_j)=prob(x_j)$. Also, define an ex-post two objects winning probability function  for $i$, $q(b_i, b_j)$ as follows.

\disparray x{
q(b_i, b_j)=\begin{cases}
1 & if\ b_i>b_j\\
1/2 & if\ [(b_{i1}>b_{j1})\ and \ (b_{i2}=b_{j2})] \ or \ [(b_{i1}=b_{j1}]) \ and \ (b_{i2}>b_{j2})] \\
1/4 & if\ b_i=b_j\\
0 & \, otherwise.
\end{cases}}

Suppose the opponent plays an arbitrary monotone pure-strategy $\beta_j(x_j)=(\beta_{j1}(x_j), \beta_{j2}(x_j))$, for all $x_j$. Then, given $\beta_j$,  bidder $i$'s interim probability for winning both objects $Q_3(b_i, \beta_j)$ is
\display y{Q_3(b_i, \beta_j)=\int_{x_j\in X^2} p(x_j)\cdot q(b_i, \beta_j(x_j)) \ dx_j}

As a result, the LHS of \eqn w can be expressed as follows.
\disparray z{&&[Q_3(\bh, \bbh,\beta_j)-Q_3(\bh, \bbl, \beta_j)] -[Q_3(\bl, \bbh, \beta_j)-Q_3(\bl, \bbl, \beta_j)] \\
&=&\int_{x_j\in X^2} p(x_j)[\{q(\bh, \bbh, \beta_j(x_j))-q(\bh, \bbl, \beta_j(x_j))\}\\&& -\{q(\bl, \bbh, \beta_j(x_j))-q(\bl, \bbl, \beta_j(x_j))\}] \ dx_j\\
&=&\int_{x_j\in X^2} p(x_j)[H(\bh, \bbl, \bbh, \beta_j(x_j))-D(\bl, \bbl, \bbh, \beta_j(x_j))] \ dx_j.
}
Note that $H(\bh, \bbl, \bbh, \beta_j(x_j))=q(\bh, \bbh, \beta_j(x_j))-q(\bh, \bbl, \beta_j(x_j))$ and $D(\bl, \bbl, \bbh, \beta_j(x_j))=q(\bl, \bbh, \beta_j(x_j))-q(\bl, \bbl, \beta_j(x_j))$. That is, $H$ represents an increase in $q$, expost winning probability for both objects,  when the agent $i$ increases the second object's bid, while bidding high for the first object. $D$ describes the same effect when $i$ bids low for the first object.\\

Now, we want to show $\forall x_j, \forall \vec{b_i}\in \{(\bl, \bbl, \bh, \bbh)| \ b_{ik}^h>b_{ik}^l, \ k=1,2\}$,  $$H(\bh, \bbl, \bbh, \beta_j(x_j))-D(\bl, \bbl, \bbh, \beta_j(x_j)) \geq 0.$$\\
Pick arbitrary $x_j$ and $\beta_j$, and let's calculate $H(\bh, \bbl, \bbh, \beta_j(x_j)) - D(\bl, \bbl, \bbh, \beta_j(x_j))$, for every possible case between $\vec{b}_i$ and $\beta_j(x_j)$.
\footnote{There are 5 main categories defined by [$\bbh, \bbl, \ and \ \beta_{j2}(x_j)$] and 5 sub categories defined by [$\bh, \bl, \ and \ \beta_{j1}(x_j)$] . The main categories are [$\bbh>\beta_{j2}(x_j) > \bbl$], [$\bbh>\beta_{j2}(x_j)=\bbl$ ], [$\bbh>\bbl >\beta_{j2}(x_j)$], [$\bbh=\beta_{j2}(x_j)>\bbl$] and [$\beta_{j2}(x_j)>\bbh>\bbl$].  Sub categories are shown in Table. 1.  } We start from the first main category, which describes the case that the second bid of the opponents is located between bidder $i$'s high and low bids for the second object. In this case, the low bid always gives value 0 for $q(\cdot, \bbl, \beta_j(x_j)).$\\

\enumerate{
\item{$\bbh>\beta_{j2}(x_j) > \bbl$\\

For $\bl$, there are 3 possibilities.\\

i) $\bl<\beta_{j1}(x_j)$\\
$$D_1^1=q(\bl, \bbh, \beta_j(x_j))-q(\bl, \bbl, \beta_j(x_j))=0-0=0\footnote{The subscript of $D$(or $H$) denotes the main category number, and the superscript describes the cases defined by first bids.}$$\\

ii) $\bl=\beta_{j1}(x_j)$
$$D_1^2=q_j(\bl, \bbh, \beta_j(x_j))-q(\bl, \bbl, \beta_j(x_j))=1/2-0=1/2$$\\

iii) $\bl>\beta_{j1}(x_j)$
$$D_1^3=q(\bl, \bbh, \beta_j(x_j))-q(\bl, \bbl, \beta_j(x_j))=1-0=1$$\\

For $\bh$, there are 3 possibilities.\\

a) $\bh<\beta_{j1}(x_j)$\\
$$H_1^1=q(\bh, \bbh, \beta_j(x_j))-q(\bh, \bbl, \beta_j(x_j))=D_1^1=0$$\\
The second equality holds, since $\bh$ plays a same role with $\bl$ in $D_1^1.$\\

b) $\bh=\beta_{j1}(x_j)$
$$H_1^2=D_1^2=1/2$$\\

c) $\bh>\beta_{j1}(x_j)$
$$H_1^3=D_1^3=1$$\\

In summary, 
\disparray A{
&&H_1^1=D_1^1=0\\
&&H_1^2=D_1^2=1/2\\
&&H_1^3=D_1^3=1\\}

\begin{table}
\centering
\caption{\texttt{$H-D$ category}}
\bigskip
\begin{tabular}{|c|c|c|c|c|}

\hline 
 &  &$H^1$  &$H^2$  &$H^3$ \tabularnewline
\hline 

 &  &  $b_1^h < \beta_{j1}(x_j)$&$b_1^h = \beta_{j1(x_j)}$  &$b_1^h > \beta_{j1}(x_j)$ \tabularnewline
\hline 
 $D_1$& $b_1^l < \beta_{j1}(x_j)$ &$H^1-D^1$  & $H^2-D^1$  & $H^3-D^1$ \tabularnewline
\hline 
$D_2$ & $b_1^l = \beta_{j1}(x_j)$ &$\oslash$  &$\oslash$  & $H^3-D^2$ \tabularnewline
\hline 
 $D_3$& $b_1^l > \beta_{j1}(x_j)$ &$\oslash$  & $\oslash$ & $H^3-D^3$  \tabularnewline
\hline 
\multicolumn{5}{c}{}\tabularnewline
\multicolumn{5}{c}  {\text{Cells with $\oslash$ are not defined,}} \\
\multicolumn{5}{c} {\text{because $[b_1^l>\beta_{j1}>b_1^h]$ contradicts to $[b_1^h>b_1^l]$.}}\tabularnewline
\end{tabular}

\end{table}

\bigskip
See the \textit{Table 1.}  We want to show that values of the cells defined are positive or zero, and it holds with the results in \eqn A. It also holds for below 4 cases.}\\

\item{$\bbh>\beta_{j2}(x_j)=\bbl$ \\
\disparray B{
&&H_2^1=D_2^1=0-0=0\\
&&H_2^2=D_2^2=1/2-1/4=1/4\\
&&H_2^3=D_2^3=1-1/2=1/2}
}

\item{$\bbh>\bbl >\beta_{j2}(x_j)$\\
\disparray C{
&&H_3^1=D_3^1=0-0=0\\
&&H_3^2=D_3^2=1/2-1/2=0\\
&&H_3^3=D_3^3=1-1=0\\}
}

\item{$\bbh=\beta_{j2}(x_j)>\bbl$\\
\disparray D{
&&H_4^1=D_4^1=0-0=0\\
&&H_4^2=D_4^2=1/4-0=1/4\\
&&H_4^3=D_4^3=1/2-0=1/2\\}
}

\item{$\beta_{j2}(x_j)>\bbh>\bbl$\\
\disparray E{
&&H_5^1=D_5^1=0\\
&&H_5^2=D_5^2=0\\
&&H_5^3=D_5^3=0\\}
}
}

We have shown that $H(\bh, \bbl, \bbh, \beta_j(x_j))-D(\bl, \bbl, \bbh, \beta_j(x_j)) \geq 0$ holds for every cases. Therefore, the inequality \eqn w holds. Note that no restriction exists for $p(x_j)$. Hence, any distribution on $X$ is allowed. 
$\blacklozenge$\\}

Therefore, by Lemma 2, the interim utility $V_i(b_i,x_i, s_{-i})$ is weakly quasi-supermodular.

\section{Conclusion}

This paper has shown the existence of monotone pure-strategy equilibrium in simultaneous first-price auctions, where the objects have synergy in bidders' utility. Under the complementarity assumptions, weak single crossing property and weak quasi supermodularity of interim utility are satisfied, so \cite{reny2011existence}'s existence theorem is applicable in this environment. \\

As a future research, it is meaningful to think about the environment in a continuum action space setting. As \cite{athey2001single} argues in the paper, it would support  a differential equation approach to verify existence conditions. She provides the extension with a continuum action space by showing ties do not occur in the game. \cite{reny2011existence} also presents the extension through a better-reply secure game and a limit of a pure-strategy of $\epsilon$ equilibria. \\



\bibliographystyle{econ}


\ifx\undefined\bysame
\newcommand{\bysame}{\hskip.3em \leavevmode\rule[.5ex]{3em}{.3pt}\hskip0.5em}
\fi


\end{document}